\documentclass[twocolumn]{aastex62}

\def\ra#1#2#3{#1$^{\rm h}$ #2$^{\rm m}$ #3$^{\rm s}$}
\def\dec#1#2#3{$#1^\circ #2' #3''$}

\begin{document}

\title{Searches After Gravitational-waves Using ARizona Observatories (SAGUARO): System Overview and First Results from Advanced LIGO/Virgo's Third Observing Run}
\newcommand{\UA}{\affiliation{Steward Observatory, University of Arizona, 933 North Cherry Avenue, Tucson, AZ 85721-0065, USA}}
\newcommand{\NU}{\affiliation{Center for Interdisciplinary Exploration and Research in Astrophysics and Department of Physics and Astronomy, \\ Northwestern University, 2145 Sheridan Road, Evanston, IL 60208-3112, USA}}
\newcommand{\UCDavis}{\affiliation{Department of Physics, University of California, 1 Shields Avenue, Davis, CA 95616-5270, USA}}
\newcommand{\Padova}{\affiliation{Department of Physics and Astronomy Galileo Galilei, University of Padova, Vicolo dell'Osservatorio, 3, I-35122 Padova, Italy}}
\newcommand{\INAF}{\affiliation{INAF Osservatorio Astronomico di Padova, Vicolo dell'Osservatorio 5, I-35122 Padova, Italy}}
\newcommand{\INAFbol}{\affiliation{INAF - Osservatorio di Astrofisica e Scienza dello Spazio - Via Piero Gobetti 93/3, I-40129 Bologna, Italy}}
\newcommand{\LPL}{\affiliation{Lunar and Planetary Lab, Department of Planetary Sciences, University of Arizona, Tucson, AZ 85721, USA}}
\newcommand{\NOAO}{\affiliation{National Optical Astronomy Observatory, 950 North Cherry Avenue, Tucson, AZ 85719, USA}}
\newcommand{\PATAU}{\affiliation{The School of Physics and Astronomy, Tel Aviv University, Tel Aviv 69978, Israel}}
\newcommand{\TTU}{\affiliation{Department of Physics and Astronomy, Texas Tech University, Box 1051, Lubbock, TX 79409-1051, USA}}
\newcommand{\OSU}{\affiliation{Department  of  Astronomy,  The  Ohio  State University,  140  W.  18th  Ave.,  Columbus,  OH43210, USA}}
\newcommand{\LBT}{\affiliation{Large Binocular Telescope Observatory, 933 North Cherry Avenue, Tucson, AZ, USA}}
\newcommand{\RMC}{\affiliation{Department of Physics and Space Science Royal Military College of Canada P.O. Box 17000, Station Forces Kingston, ON K7K 7B4, Canada}}
\newcommand{\ASU}{\affiliation{School of Earth and Space Exploration, Arizona State University, Tempe, AZ 85287, USA}}
\newcommand{\MMT}{\affiliation{MMT Observatory, PO Box 210065, University of Arizona, Tucson, AZ 85721-0065, USA}}
\newcommand{\NAU}{\affiliation{Department of Physics and Astronomy, Northern Arizona University, P.O. Box 6010, Flagstaff, AZ 86011, USA}}
\newcommand{\UAOptSci}{\affiliation{College of Optical Sciences, University of Arizona, 1630 E University Blvd, Tucson, AZ 85719, USA}}
\newcommand{\UNC}{\affiliation{Department of Physics and Astronomy, University of North Carolina at Chapel Hill, Chapel Hill, NC 27599, USA}}
\newcommand{\MSU}{\affiliation{Center for Data Intensive and Time Domain Astronomy, Department  of  Physics  and  Astronomy,  Michigan  State  University,East Lansing, MI 48824, USA}}
\newcommand{\UCSC}{\affiliation{Department of Astronomy and Astrophysics, University of California, Santa Cruz, CA 95064, USA}}
\newcommand{\STScI}{\affiliation{Space Telescope Science Institute, 3700 San Martin Drive, Baltimore, MD 21218, USA}}
\newcommand{\Brandeis}{\affiliation{Department of Physics, Brandeis University, Waltham, MA 02453, USA}}
\newcommand{\LCO}{\affiliation{Las Cumbres Observatory, 6740 Cortona Drive, Suite 102, Goleta, CA 93117-5575, USA}}
\newcommand{\UToronto}{\affiliation{Department of Astronomy and Astrophysics, University of Toronto, 50 St. George Street, Toronto, Ontario, M5S 3H4 Canada}}
\newcommand{\NotreDame}{\affiliation{Department of Physics, University of Notre Dame, Notre Dame, IN 46556, USA}}
\newcommand{\UMN}{\affiliation{College of Science \& Engineering, Minnesota Institute for Astrophysics, University of Minnesota, 115 Union St. SE, Minneapolis, MN 55455, USA}}
\newcommand{\UT}{\affiliation{Department of Astronomy, University of Texas at Austin, Austin, TX 78712, USA}}
\newcommand{\JHU}{\affiliation{The Johns Hopkins University, Baltimore, MD 21218, USA}}
\newcommand{\VAT}{\affiliation{Vatican Observatory, 00120 Citt\`{a} del Vaticano, Vatican City State  }}
\newcommand{\HF}{\affiliation{Hubble Fellow}}
\newcommand{\Carnegie}{\affiliation{The Observatories of the Carnegie Institution for Science, 813 Santa Barbara St., Pasadena, CA 91101, USA}}

\author[0000-0001-9589-3793]{M.~J. Lundquist}
\UA

\author[0000-0001-8340-3486]{K.~Paterson}
\NU

\author[0000-0002-7374-935X]{W.~Fong}
\NU

\author[0000-0003-4102-380X]{D.~J. Sand}
\UA

\author[0000-0003-0123-0062]{J.~E. Andrews}
\UA

\author[0000-0003-4702-7561]{I.~Shivaei}
\UA\HF

\author{P.~N. Daly}
\UA

\author[0000-0001-8818-0795]{S.~Valenti}
\UCDavis

\author[0000-0002-2898-6532]{S.~Yang}
\UCDavis\Padova\INAF

\author{E.~Christensen}
\LPL

\author[0000-0002-2575-2618]{A.~R. Gibbs}
\LPL

\author{F.~Shelly}
\LPL

\author{S.~Wyatt}
\UA

\author{O.~Kuhn}
\LBT



\author[0000-0002-1546-9763]{R.~C. Amaro}
\UA

\author[0000-0001-7090-4898]{I. Arcavi}
\PATAU

\author[0000-0002-2517-6446]{P. Behroozi}
\UA


\author{N. Butler}
\ASU

\author[0000-0002-8400-3705]{L. Chomiuk}
\MSU

\author[0000-0003-3433-1492]{A.~Corsi}
\TTU

\author{M.~R. Drout}
\UToronto\Carnegie

\author{E. Egami}
\UA

\author{X. Fan}
\UA

\author{R.~J. Foley}
\UCSC

\author{B. Frye}
\UA

\author{P. Gabor}
\VAT


\author{E.~M. Green}
\UA

\author[0000-0001-9920-6057]{C.~J. Grier}
\UA

\author[0000-0001-9136-929X]{F. Guzman}
\UAOptSci\UA

\author{E. Hamden}
\UA

\author[0000-0003-4253-656X]{D. A. Howell}
\LCO

\author[0000-0002-1578-6582]{B. T. Jannuzi}
\UA

\author[0000-0003-3142-997X]{P. Kelly}
\UMN




\author{P. Milne}
\UA

\author{M. Moe}
\UA

\author[0000-0002-2028-9329]{A.~Nugent}
\NU

\author{E. Olszewski}
\UA


\author[0000-0002-8691-7666]{E. Palazzi}
\INAFbol

\author{V. Paschalidis}
\UA

\author{D. Psaltis}
\UA

\author{D. Reichart}
\UNC

\author[0000-0002-4410-5387]{A. Rest}
\STScI\JHU

\author[0000-0002-8860-6538]{A. Rossi}
\INAFbol

\author[0000-0001-9915-8147]{G.~Schroeder}
\NU

\author{P. S. Smith}
\UA

\author{N.~Smith}
\UA


\author[0000-0002-0956-7949]{K. Spekkens}
\RMC

\author[0000-0002-1468-9668]{J. Strader}
\MSU

\author{D. P. Stark}
\UA

\author{D. Trilling}
\NAU

\author{C. Veillet}
\UA\LBT

\author{M. Wagner}
\OSU\LBT

\author[0000-0001-6065-7483]{B. Weiner}
\UA\MMT

\author[0000-0003-1349-6538]{J. C. Wheeler}
\UT

\author{G.~G.~Williams}
\UA\MMT

\author[0000-0001-6047-8469]{A. Zabludoff}
\UA

\begin{abstract}

We present Searches After Gravitational-waves Using ARizona Observatories (SAGUARO), a comprehensive effort dedicated to the discovery and characterization of optical counterparts to gravitational wave (GW) events. SAGUARO utilizes ground-based facilities ranging from 1.5m to 10m in diameter, located primarily in the Northern Hemisphere. We provide an overview of SAGUARO's telescopic resources, pipeline for transient detection, and database for candidate visualization. We describe SAGUARO's discovery component, which utilizes the $5$~deg$^2$ field-of-view optical imager on the Mt. Lemmon 1.5m telescope, reaching limits of $\approx 21.3$~AB mag while rapidly tiling large areas. We also describe the follow-up component of SAGUARO, used for rapid vetting and monitoring of optical candidates.
With the onset of Advanced LIGO/Virgo's third observing run, we present results from the first three SAGUARO searches following the GW events S190408an, S190425z and S190426c, which serve as a valuable proof-of-concept of SAGUARO. We triggered and searched 15, 60 and 60 deg$^{2}$ respectively, 17.6, 1.4 and 41.8 hrs after the initial GW alerts. We covered 7.8, 3.0 and 5.1\% of the total probability within the GW event localizations, reaching 3$\sigma$ limits of 19.8, 21.3 and 20.8 AB mag, respectively. Although no viable counterparts associated with these events were found, we recovered 6 known transients and ruled out 5 potential candidates. We also present Large Binocular Telescope spectroscopy of PS19eq/SN2019ebq, a promising kilonova candidate that was later determined to be a supernova. With the ability to tile large areas and conduct detailed follow-up, SAGUARO represents a significant addition to GW counterpart searches.

\end{abstract}

\section{Introduction} \label{sec:intro}

The onset of the advanced era of gravitational wave (GW) detectors has heralded a new era of discovery. The Advanced Laser Interferometer Gravitational-wave Observatory (\citealt{lvc_ligo}) and Advanced Virgo (\citealt{lvc_virgo}; hereafter termed ``LVC'') have discovered a total of 11~GW events in their first two observing runs (O1-O2; \citealt{lvc_O1O2}), including ten~binary black holes (BBH) mergers as well as the first binary neutron star (BNS) merger, GW170817 \citep{lvc_gw170817}. However, identifying an electromagnetic (EM) counterpart presents an observational challenge, as GW events thus far have been localized to $\approx 16-1650$~deg$^2$ (90\% confidence; \citealt{lvc_O1O2}), often requiring wide-field imagers to cover a meaningful fraction of the localization regions for photometric discovery.

Two primary approaches have been taken to identify optical counterparts to GW events: ``galaxy-targeted'' searches which focus on plausible galaxies within the GW localization regions (e.g., \citealt{Gehrels16}), and wide-field searches which cover more substantial areas on the sky and are relatively agnostic to the distribution of stellar mass within a GW localization. GW170817 was localized to $\approx$30~deg$^2$ from the GW signal alone \citep{lvc_gw170817} at the time of the initial counterpart searches, and both wide-field and galaxy-targeted strategies proved fruitful for the discovery of the optical counterpart (e.g., \citealt{Arcavi17_2,Coulter17,Lipunov17,Tanvir17,Soares17,Valenti2017}). 
Looking forward, the median localization of BNS mergers for the third LVC observing run (``O3'') is predicted to be $\approx 120-180$~deg$^2$ for events detected by all three LIGO/Virgo detectors \citep{lvc_loc}. Moreover, most BNS mergers are expected to be detected at distances of $\gtrsim\!100$~Mpc, where galaxy catalogs are incomplete \citep{GWGC,glade}, motivating dedicated wide-field searches to discover optical counterparts.

In this paper, we describe a telescope network brought online in LVC's O3 dedicated to optical counterpart discovery and follow-up of GW events: Searches After Gravitational-waves Using ARizona Observatories (SAGUARO). In Section~\ref{sec:saguaro} we provide an overview of SAGUARO's scope and telescopic resources. In Section~\ref{sec:CSS} we describe our wide-field photometric counterpart search, automated pipeline for transient discovery, database for candidate visualization, and current status. In Section~\ref{sec:followup} we describe our resources and methods for spectroscopic classification of candidates and concentrated follow-up of true EM counterparts. In Section~\ref{sec:events} we present results from the first three SAGUARO searches following the GW events S190408an, S190425z and S190426c as proof-of-concept studies. Finally, in Section~\ref{sec:summary} we summarize and discuss future prospects.

Unless otherwise stated, all magnitudes reported here are in {\it Gaia} $G$-band and are converted to the AB system via $m_{\rm AB}=m_{\rm Gaia}+0.125$ \citep{Maiz18}.

\section{SAGUARO Overview and Scientific Scope  }\label{sec:saguaro}

The SAGUARO GW follow-up program has two distinct but intertwined components: 1) a wide-field optical search for EM counterparts utilizing the Steward Observatory 1.5m Mt.~Lemmon telescope and its 5 deg$^2$ imager; 2) a comprehensive optical and near-infrared (NIR) follow-up program composed largely of Steward Observatory telescopes, but also including a few programs outside of Arizona.  We detail these components in Sections~\ref{sec:CSS} and \ref{sec:followup}, respectively.

The search component of SAGUARO utilizes the existing infrastructure and personnel of the Catalina Sky Survey \citep[CSS;][]{CSS} to respond in real time to GW events of interest.  Given the 5 deg$^2$ field of view of the imager, we employ a wide-field search strategy that directly tiles the GW localization region, similar to that of other groups with access to wide-field facilities \citep[e.g.][]{Smartt16,Kasliwal16,GWem,Soares17,Goldstein19}, rather than a galaxy-targeted approach \citep[e.g.][]{Gehrels16,Arcavi17,DLT40_O2}.

Once viable EM counterpart candidates are discovered and vetted, SAGUARO has direct access to several optical/NIR follow-up telescopes with apertures ranging from 1.5--10-m, and Target-of-Opportunity (ToO) programs to enable rapid imaging and spectroscopic follow-up.  We detail these facilities further in Section~\ref{sec:followup} and demonstrate that detecting kilonova emission out to $\sim$200 Mpc and beyond is feasible within the SAGUARO framework. From these measurements we can constrain properties such as composition, mass and velocity of ejected material (e.g., \citealt{Pian2017}, \citealt{Barnes2013}, \citealt{Metzger17}, \citealt{Chornock2017}). Of particular interest is the direct constraint on the production of heavy elements that the ejected mass, together with the merger rates, can provide.

The final component of SAGUARO is to characterize the galactic environments of compact object mergers, which has proved fruitful in constraining the formation histories of the progenitor systems of GW170817 \citep{gw170817progenitor,Blanchard17,Levan17,Pan17} and cosmological short-duration $\gamma$-ray bursts (SGRBs) \citep{Fong13}. For a given GW event, once a single optical counterpart is found, SAGUARO will utilize multi-band optical/NIR photometry and spectroscopy of the host galaxy to enable inferences on the global stellar population properties (e.g., stellar mass, star formation rate, stellar population age, star formation history), and spatially-resolved spectroscopy, which can constrain properties of the preferred merger sites of NS binaries and provide filtering for potential hosts in galaxy-targeted searches.

SAGUARO is active for O3 and beyond in order to address outstanding questions concerning the physics and emission mechanisms of these novel, multi-messenger cosmic explosions \citep[e.g.][for a review]{Metzger17}.

\section{Optical Counterpart Search \& Data Flow}  \label{sec:CSS}

The primary search capability for SAGUARO utilizes the wide-field imaging of CSS, a near-Earth object (NEO) and potentially hazardous asteroids (PHA) discovery and characterization program.  We briefly describe the relevant aspects here.

While CSS utilizes several telescopes, we are currently using the Steward Observatory 1.5-m Mt. Lemmon telescope for our EM counterpart discovery program.  The telescope is equipped with a prime focus imager and a 10.5K$\times$10.5K CCD (0.77\arcsec\ per pixel), resulting in a 5 deg$^2$ field of view (FOV).  It is operated with 2$\times$2 binning for an effective plate scale of 1.54\arcsec\ per pixel.
In order to discover NEOs and PHAs, CSS visits fields four times in a $\sim$30 min time span to identify moving objects.  With 30 second exposures and typical overheads of several seconds, CSS observes 12 fields, covering 60 deg$^2$, in such a 30 min set.  All images are taken without a filter, and the typical 3$\sigma$ image depth of a 4$\times$30 sec set of median-combined images is $G$$\approx$21.3 mag, calibrated to {\it Gaia} DR2 \citep{Gaiadr2}.

CSS observes fixed fields on the sky, between $-$25~deg and +60~deg in Declination, while avoiding crowded regions in the Galactic plane (see Figure~\ref{fig:CSSfields}).  The team observes $\sim$24 nights per month, avoiding the period around full moon (our GW counterpart search is also not operable during this time period).  Once an appropriate GW is announced (see trigger criterion below), the CSS team will observe a 60 (or 120) deg$^2$ set of images within the localization region, taking the same sequence of 4 images as is done for their standard NEO search.

In the following subsections we discuss the logistics of triggering our CSS search when a GW event is announced, and our real time pipeline for difference imaging and transient detection.  We end the section by discussing the current status of this wide-field GW search program.

\subsection{Triggering CSS}

The SAGUARO software suite ingests the VOEvents distributed by the NASA Gamma-Ray Coordinates Network (GCN)\footnote{\url{https://gcn.gsfc.nasa.gov/lvc.html}}  system in real time.  These alerts are employed by the LVC for disseminating GW event information, including the {\sc HEALPIX} localization maps with distance constraints \citep[e.g.][]{Singer16,SingerPrice16}.  The alert contents are described in the LIGO/Virgo Public Alerts User Guide\footnote{\url{https://emfollow.docs.ligo.org/userguide/index.html}} and contain the classification probabilities for each GW event, which are split into five categories \citep[see][for more details]{Kapadia19}: 1) terrestrial (not of astrophysical origin), 2) BNS (both components are neutron stars, 1 $<$ M $<$ 3 $\mathrm{M}_{\odot}$), 3) MassGap (any component has a mass in the gap between neutron stars and black holes, 3 $<$ M $<$ 5 $\mathrm{M}_{\odot}$), 4) NSBH (one component is a neutron star, 1 $<$ M $<$ 3 $\mathrm{M}_{\odot}$ and one is a black hole M $>$ 5 $\mathrm{M}_{\odot}$ ), and 5) BBH (both components are black holes, M $>$ 5 $\mathrm{M}_{\odot}$).  The alerts also contain two parameters, HasNS and HasRemnant, that indicate the likelihood that the event produces EM emission.  HasNS indicates the probability that one of the components was a neutron star and HasRemnant indicates the probability that some material remained outside the final remnant compact object, as calculated by the \citet{Foucart18} model.

Once an alert is received, our CSS search is automatically triggered if the classification parameters exceed our limits for triggering, the false alarm rate is better than our requirement, and the target is observable by CSS.  We require that the classification for the GW event have a combined BNS, NSBH, and MassGap probability greater than 20\% and a false alarm rate $<$ 12 yr$^{-1}$ in order to trigger.  The GW event is determined to be observable if any of the CSS fields within the 90\% probability region meet the following constraints: 1) airmass $<$ 2.5, 2) the projected distance on the sky to the Moon, $d_{moon}$, is governed by $d_{moon} > (42\times \theta_{illum} + 3)$ deg, such that the $d_{moon}$ limit increases with the moon illumination $\theta_{illum}$ (represented by a fractional number going from 0 for new to 1 for full), 3) sun altitude $<$ -12 deg.  The exception to this was the first event of the LIGO/Virgo O3 run, S190408an, which was used as a full system test even though it was a clear BBH (see Section~\ref{sec:events}).  

Once it is determined that a GW event meets our criteria for triggering, the SAGUARO software automatically inserts a selection of up to 12 fields (60 deg$^2$) into the CSS observing queue with the option of manually triggering an additional 12 fields.  These fields are selected to cover the highest probability regions that are observable and for which we have template images.  They are given a higher priority level that allows them to be observed immediately after the current CSS sequence is finished.

\subsection{CSS Discovery Data }
Once all 4 images of a field have been taken, basic processing including bias subtraction and flat fielding are performed.  Astrometric and photometric calibrations are then done with SCAMP \citep{scamp,scamp2} using {\it Gaia} DR2,  resulting in a typical standard deviation of $\sim$0.11\arcsec\ and $\sim$0.19 mag for the astrometry and zero point, respectively.

A script watching for new data waits for all 4 images of a field to arrive before creating a median for that field using SWarp \citep{swarp}. During the median creation, each image is background subtracted to remove background variations before median combining. The background value is retained and used for scaling purposes on the median.  The creation of medians allows for cleaner images on which we can search for transients, and removes artifacts such as cosmic rays that only appear on a single image. As an image set for a particular field can be separated by $\sim30$ min, there exists the possibility that not all four images for each field will be observed, due to weather, moon or sun constraints -- we therefore have built in a time limit for our median creation algorithm to create medians with the available images if all 4 are not received.  
A real-time data processing pipeline, described in more detail in Section \ref{sec:pipeline}, then processes these median images for transient detection.

\subsection{Creation of CSS Templates} \label{sec:template}
\begin{figure*}
    \centering
    \includegraphics[width=\textwidth]{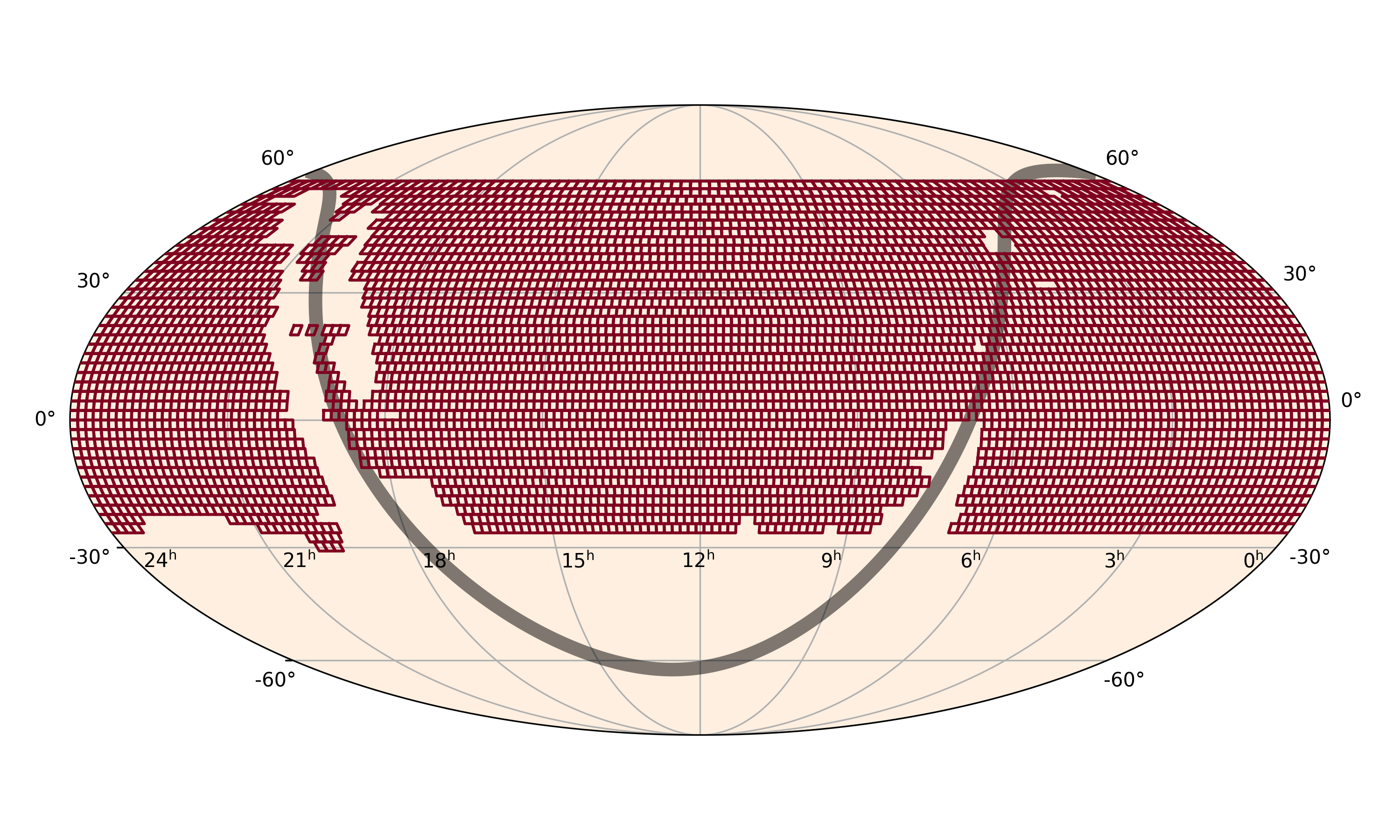}
    \vspace{-0.5in}
    \caption{All-sky projection of the CSS survey coverage.  Each square represents a 5.0 deg$^{2}$ CSS field for which we have a template image.  We have templates for 5069 fields covering 25345 deg$^2$.  The Galactic plane is indicated by the shaded line.  Crowded regions near the Galactic plane are avoided, resulting in the two bands of missing templates around RA $\approx$ 7h and  $\approx$ 19h.
    }
    \label{fig:CSSfields}
\end{figure*}

One of the primary advantages of SAGUARO's transient discovery is the access to nearly three years of archival data with the current instrumentation, which enable the production of deep templates for image subtraction.  
We created templates that cover the entire CSS footprint of 25345 deg$^2$ (see Figure~\ref{fig:CSSfields}) with a median of $\sim$60 individual 30-sec images contributing to each template after a series of quality cuts. First, we discard images which have $<2000$~detected point sources. This is done to ensure high astrometric precision for the templates. Second, we do not include images which have a sky brightness of $<20$~mag/arcsec$^2$. This cut was chosen based on the distribution of measured sky brightness values, and serves to exclude poor quality data taken in adverse conditions (i.e. clouds or bright moon). Applying these cuts restricts the template creation to high-quality, deep images with a median $3\sigma$ limiting mag of $23.0$~mag. Approximately 27\% of the archival data was not included in the template creation. At the two extremes, 5\% of the templates were created with more than 90 images per field while 9\% of the templates were only observed once and have $\leq$ four images per field. For those fields with poor coverage (e.g. at very high declination or close to the Galactic plane), no quality cuts were applied and all available images were used to create the template.

\subsection{Transient Pipeline} \label{sec:pipeline}

SAGUARO makes use of a data processing pipeline written in Python (Paterson et al., in prep will discuss the pipeline in detail) and an implementation of the image subtraction method called  ZOGY\footnote{\url{https://github.com/pmvreeswijk/ZOGY}} \citep{ZOGY}.

The data processing pipeline controls the flow of the data, submitting images for image subtraction and creating detailed logs for each image. The pipeline also creates a mask for each image by identifying saturated stars and bad pixels. This allows regions with poor subtractions caused by bad pixels and saturated stars to be ignored during the transient detection.

Image subtraction is performed using the new median images and the templates discussed in Section \ref{sec:template}. The difference image created by the image subtraction is then converted into a significance image, where the value of each pixel is represented by its significance (i.e. a pixel with value 5 will have a  signal-to-noise ratio (SNR) of 5). Corrections, such as those for astrometric errors, are applied to the significance image to produce a corrected significance or ``Scorr'' image (see \citealt{ZOGY} for a detailed description on the products produced). As the significance or SNR is encoded directly in the Scorr image, it provides a direct way to find detections above a set threshold. For SAGUARO, we set the detection threshold on the subtracted images to 5$\sigma$. Thus, sources with a significance or SNR$>5$ on the Scorr image will be flagged as transient candidates. A flux for each detection is obtained through point spread function (PSF) photometry. The zero-point, calculated through SCAMP \citep{scamp,scamp2} using {\it Gaia} DR2, is then used to convert this flux to magnitudes in {\it Gaia} $G$-band (roughly covering 330 nm to 1050 nm; see \citealt{Weiler2018}). With the focus on transients, we crossmatch detections against stellar sources in the templates to remove variable stars and poor subtractions associated with stars close to saturation. At present, SExtractor's CLASS\_STAR parameter is used to filter stellar sources (defined as CLASS\_STAR $>$ 0.5). The remaining candidates are then loaded into the database for visualization and vetting.

\subsection{Database and Candidate Visualization}
 
For vetting in real time, detailed information for each transient candidate is stored in a PostgreSQL database and postage stamp images are saved to disk.  A Flask webserver allows visual inspection of candidates as they come in, and accommodates  queries based on date, field ID and detection threshold.    Candidates are sortable by SNR or machine learning score (which gives the likelihood of the transients being real) to promote the most likely candidates.

Each candidate is automatically cross-matched against known moving objects from the Minor Planet Center (MPC\footnote{\url{https://www.minorplanetcenter.net/iau/mpc.html}}) and known transients from the Transient Name Server (TNS\footnote{\url{https://wis-tns.weizmann.ac.il/}}).  We also search for previous detections from the Zwicky Transient Facility (ZTF; \citealt{ZTF}) and cross-match against galaxies from the Galaxy List for the Advanced Detector Era (GLADE; \citealt{glade}) catalog within the localization volume using a simple broker, the Steward Alerts for Science System (SASSy\footnote{\url{http://sassy.as.arizona.edu/sassy/ztf/}}).    
This allows for lists of viable candidates to be quickly disseminated to the community through GCN notices within a few hours of taking the data.

\begin{figure*}
\begin{center}
\includegraphics[width=0.49\textwidth]{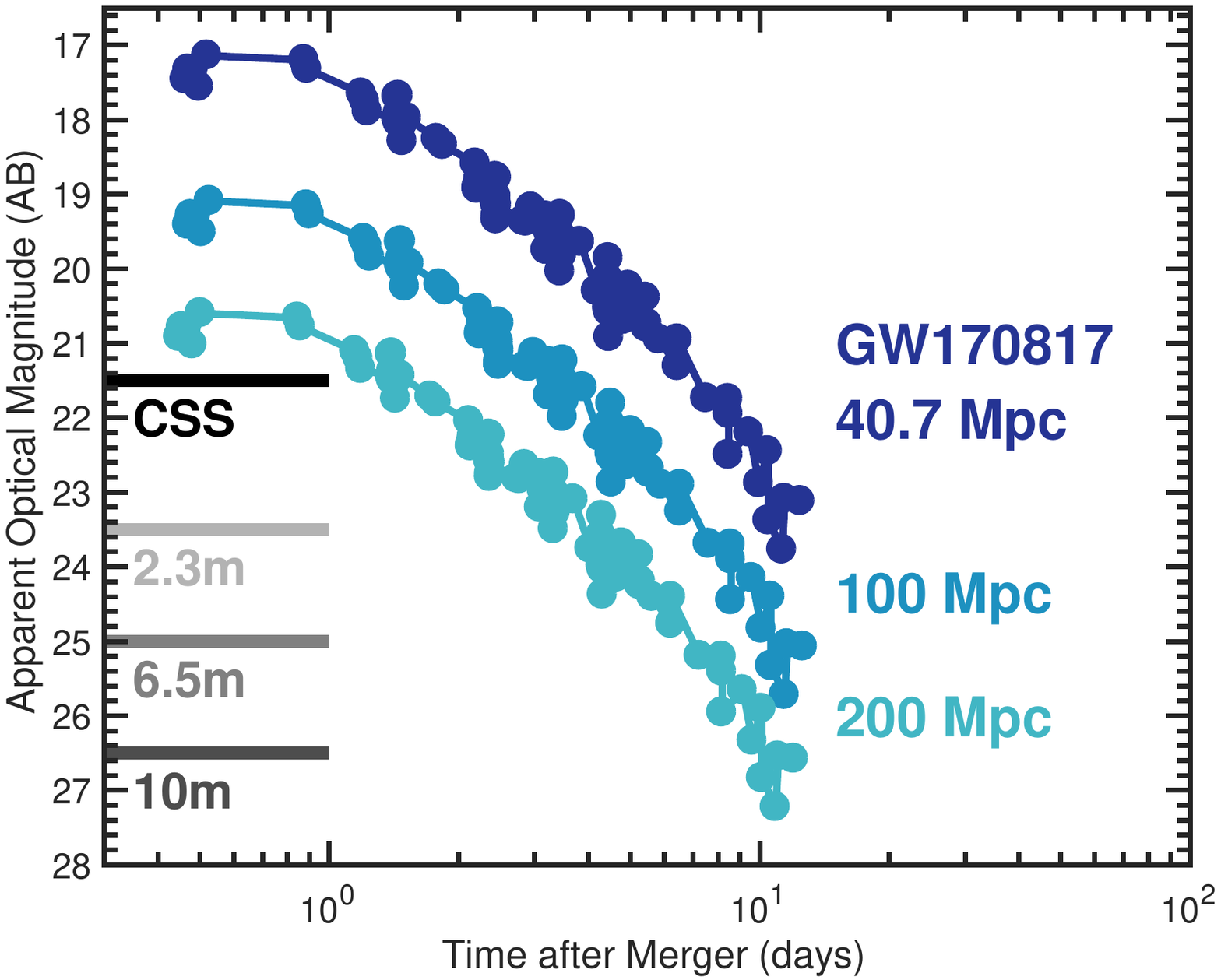}
\includegraphics[width=0.49\textwidth]{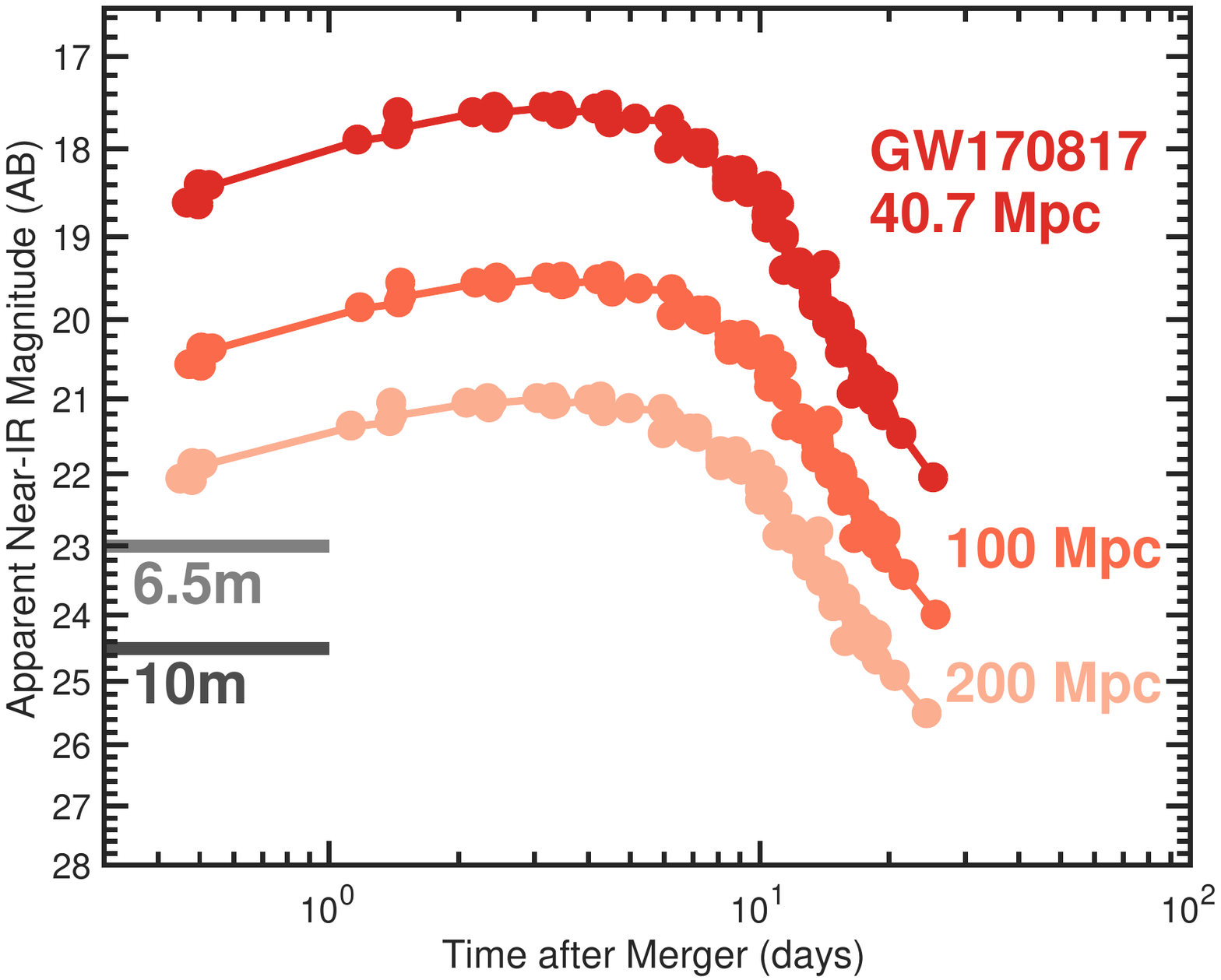}
\end{center}
\vspace{-0.1in}
\caption{Optical $r$-band (left) and NIR $K$-band (right) light curves of the $r$-process kilonova following the NS-NS merger GW170817 (compiled in \citealt{Villar17}), at the distance of the host galaxy, $\approx\!40.7$~Mpc \citep{Cantiello18}. Also shown are the light curves of GW170817, shifted to 100~Mpc, and 200 Mpc (the expected range for a face-on BNS merger at design sensitivity). The typical limit of our initial CSS search, $G\approx 21.3$~mag, is denoted by the horizontal black line. Also shown are typical $3\sigma$ limits for 30~min on-source representing the wide range of photometric follow-up resources within SAGUARO (horizontal gray-scale lines). A comparison of the scaled GW170817 light curves to depths achieved by facilities in SAGUARO demonstrate that we will be able to efficiently detect and photometrically characterize kilonovae of similar luminosity to GW170817 to $200$~Mpc in both the optical and NIR bands.}
\label{fig:gw170817}
\end{figure*}

\subsection{Current Status}

As we show in Section~\ref{sec:events}, the EM counterpart search portion of SAGUARO is functioning and responding to LVC events in O3.  We have also been ingesting all 1.5m Mt. Lemmon telescope data during routine CSS operations in real time and searching for transients \citep[e.g.][]{saguaro19a}, independent of the main GW follow-up program.  At the moment, the primary purpose of this wide-area search is to stress test our GW search pipeline, and gather a sufficient training set to improve our machine learning algorithm for transient vetting.

SAGUARO's wide-field CSS search reaches a limiting magnitude of $G\approx 21.3$~mag. In Figure~\ref{fig:gw170817}, we show the optical $r$-band and NIR $K$-band light curves for AT2017gfo, the optical counterpart of GW170817, (compiled in \citealt{Villar17}; data from \citealt{Andreoni17,Arcavi17_2,Cowperthwaite17,Coulter17,Diaz17,Drout17,Kasliwal17,Lipunov17,Pian17,Pozanenko17,Smartt17,Tanvir17,Troja17,Utsumi17,Valenti2017}) for three distances: $\approx 40.7$~Mpc, the distance of GW170817 \citep{Cantiello18}, 100~Mpc, the typical expected distance for a BNS merger for LVC's O3, and 200~Mpc, the expected range for a face-on BNS merger at design sensitivity \citep{lvc_loc}. Matched to SAGUARO's current discovery limits, our optical search maintains sensitivity to kilonovae of comparable luminosity to GW170817 out to 200~Mpc at $\delta t \lesssim 1$~day. Once a promising kilonova candidate is identified, SAGUARO's  follow-up efforts on larger aperture facilities (described in subsequent sections) will be employed to track the counterpart's temporal evolution. Indeed, it is the combined discovery and follow-up endeavor, the template coverage, and the extension to NIR wavelengths, that give SAGUARO an advantage in GW counterpart efforts.

In the Northern Hemisphere, there are only a few optical $>1$~deg$^{2}$ wide-field efforts underway to search for optical counterparts to GW events. This includes efforts by Pan-STARRS \citep{PS1}, ZTF \citep{ZTF}, and ATLAS \citep{ATLAS}. All three facilities are conducting optical counterpart searches following GW events in addition to their primary surveys. Among these searches, the SAGUARO discovery effort is most comparable to Pan-STARRs, which has a 1.8m aperture and $7$~deg$^2$ FOV. The discovery components of ZTF and ATLAS are $\approx 1$-$2$~mag less sensitive, but both possess a FOV that is a factor of $\approx 6$-$7$~times larger than SAGUARO/CSS.

\section{EM Counterpart Follow up} \label{sec:followup}

The primary facilities used by SAGUARO are those of Steward Observatory, which has significant access to optical and NIR telescopes, most of which can contribute to the follow-up of EM counterparts.  A rapid ToO program is in place to ensure timely observations in all counterpart scenarios, and our team will have access to telescopes of all relevant apertures on a given night.  Given the excitement of the burgeoning field of multi-messenger astronomy, many of the primary telescope users in the Steward community have joined the SAGUARO follow-up proposal to help facilitate rapid and persistent follow-up of any EM counterparts identified in O3.  In addition to Steward resources, we also have Keck telescope access for ToO follow-up.

\subsection{Spectroscopic Candidate Vetting}

When GW localizations cover $\sim$100-1000s deg$^2$, potentially dozens of candidate counterparts will be uncovered, as was the case for the first two potential neutron star mergers of O3, S190425z and S190426c \citep[see, e.g. the summary of these events in][]{Hosseinzadeh19}.  While this large number of candidates can be winnowed down by making cuts on the transients' age, color and association with galaxies at the appropriate distance, ultimately spectroscopic vetting must be done to uncover the true EM counterpart.  

SAGUARO is committed to spectroscopic candidate vetting for very promising targets that are confirmed to have no detection before the GW event. Candidates from all programs, including our own Mt. Lemmon 1.5-m search, will be considered equally. When distance information is available, we prioritize targets that are localized near catalogued galaxies (e.g., \citealt{glade}) with distances consistent with that inferred from the GW signal.   
Candidate vetting is a dynamic process that benefits from real time access to large aperture telescopes, as was the case for our team's vetting of PS19qp/SN2019ebq \citep{PS19qp_discover}, which an initial spectrum suggested was consistent with a kilonova at the distance to S190425z \citep{PS19qp_pessto}; we detail these observations further in Section~\ref{sec:S190425z}.

\subsection{Photometric and Spectroscopic Monitoring}

Once a true GW counterpart is discovered, SAGUARO will spring into action to collect high cadence optical and NIR light curves as well as spectral sequences while the transient is accessible to ground-based observatories.

First, in the small to medium aperture range are the 1.5m Kuiper, 1.8m VATT and 2.3m Bok telescopes, all of which are based in Southern Arizona.  All three telescopes have imagers that can gather data on kilonovae, VATT and Bok have spectrographs,  and Kuiper and Bok  have access to SPOL, an imager/spectropolarimeter \citep{SPOL}.  For instance, in $\sim$1 hour exposure times the Bok B\&C spectrograph can get high signal-to-noise spectra down to $\sim$19th mag; GW170817/AT2017gfo was brighter than 19th mag for $\sim$2 days.  Similarly, the 90Prime 1 deg$^2$ camera \citep{90Prime} on the Bok telescope can image down to $\sim$24th mag in $\sim$30 min exposures, facilitating kilonova optical light curve follow-up for a week or more in the nearest events (Figure~\ref{fig:gw170817}). 

SAGUARO also has access to large aperture facilities: the 6.5m MMT, the twin 6.5m Magellan telescopes, the 2$\times$8.4m Large Binocular Telescope, and the two Keck 10m telescopes.  The spectral sequences that these facilities can provide will lend insight into the emission mechanisms of neutron star mergers and r-process element production.
Additionally, all of these large aperture facilities will be used to obtain late-time light curves once any counterpart is too faint for spectroscopy, potentially down to $\sim$26.5 mag in the optical and $\sim$24.5 mag in the NIR.  These same large aperture facilities will be used to study the host properties of the EM counterparts to GW events, as was briefly discussed in Section~\ref{sec:saguaro}.

\section{SAGUARO Observations of GW Events} \label{sec:events}

The SAGUARO Mt. Lemmon 1.5-m search program has been activated three times thus far during O3, including the first event of the run, S190408an, a clear BBH merger, and two events that likely had a neutron star, S190425z and S190426c.  A summary of the followup observations can be seen in Table~\ref{tab:obs}.  In Figures~\ref{fig:S190408an}, ~\ref{fig:S190425z} and ~\ref{fig:S190426c}, respectively, we present each localization and our SAGUARO pointings and further describe each trigger below.

\begin{deluxetable*}{lcccccccccc}[t]
\tabletypesize{\small}
\tablecolumns{11}
\tablewidth{0pc}
\tablecaption{Summary of SAGUARO Follow-up
\label{tab:obs}}
\tablehead{
\colhead{}			 &
\colhead{}			 &
\colhead{}			 &
\colhead{}			 &
\multicolumn{3}{c}{Area covered} &
\colhead{}			 &
\multicolumn{3}{c}{Probability covered} \\
\cline{5-7}\cline{9-11} 	\\
\colhead {Event}	 &
\colhead {Type$^\dagger$}	 &
\colhead {$\delta t$}	 &	
\colhead {$3\sigma$ Limit$^{\ddagger}$} &
\colhead {50\%}  &		
\colhead {90\%}  &
\colhead {Total}  &
\colhead{}			 &
\colhead {50\%}  &		
\colhead {90\%}  &
\colhead {Total}   \\
\colhead {}		    &
\colhead {}		    &
\colhead {(hr)}  &
\colhead {(AB Mag)} &
\colhead {(deg$^2$)}	&
\colhead {(deg$^2$)}	&
\colhead {(deg$^2$)}	&
\colhead {}		    &
\colhead {(\%)}		    &
\colhead {(\%)}		    &
\colhead {(\%)}			
}
\startdata
S190408an & BBH & 17.56 & 19.8 & 8.6 & 13.5 & 15.0 &  & 13.8 & 8.6 & 7.8 \\
S190425z & BNS & 1.37 & 21.3 & 58.5 & 60.0 & 60.0 & & 6.0 & 3.4 & 3.0 \\
S190426c & NSBH & 41.76 & 20.8 & 13.4 & 58.9 & 60.0 & & 4.3 & 5.6 & 5.1 
\enddata
\tablecomments{Magnitudes reported here are uncorrected for Galactic extinction and are reported in $G$-band.\\
The probability covered refers to the percent of the probability of the 50\%, 90\%, and total localizations that were covered by these observations.\\
$^\ddagger$ $3\sigma$ limiting magnitude calculated from CSS~images. \\
$^\dagger$ Most likely classification based on GW probabilities \citep{Kapadia19} \\
}
\end{deluxetable*}

\subsection{S190408an}
\begin{figure*}
    \centering
    \includegraphics[trim=2.5cm 0cm 3cm 0cm,width=8.5cm]{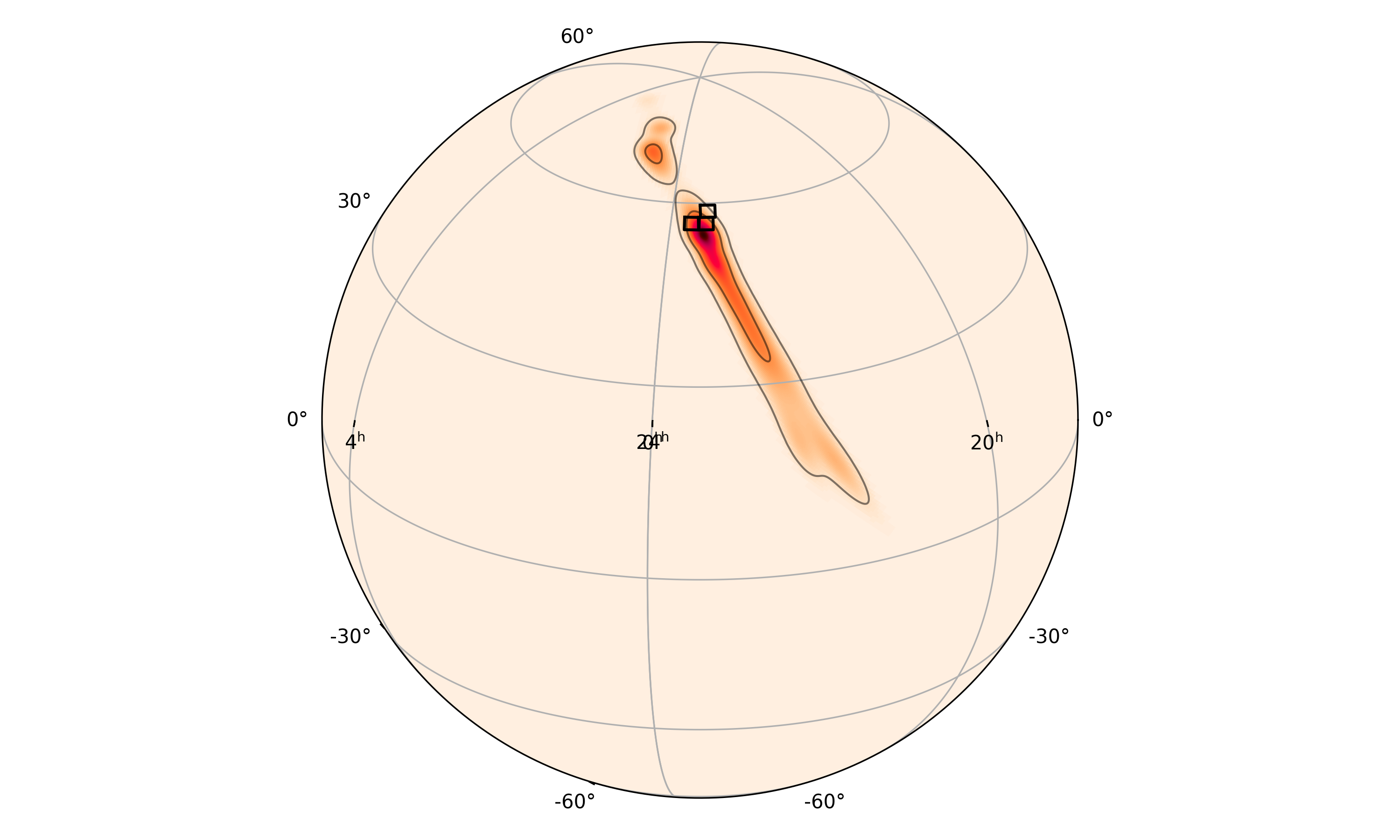}
    \includegraphics[trim=3cm 0cm 2.5cm 0cm,width=7cm, width=8.5cm]{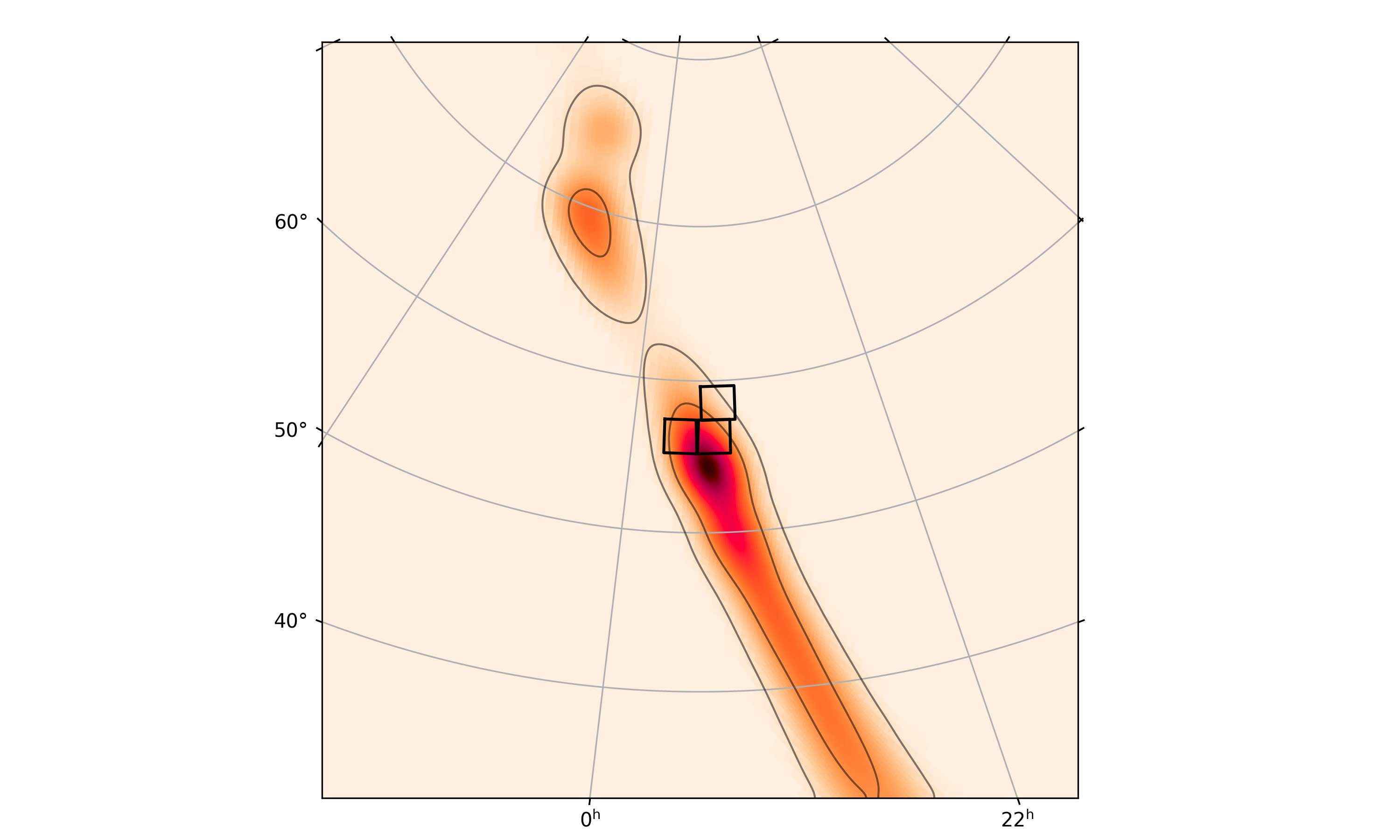}
    \caption{GW localization of S190408an overlaid with the three CSS fields that were triggered; for this event, we were limited by the number of CSS template fields available near the Galactic plane.  The 90\% localization covered 387 deg$^2$ at a distance of $1473 \pm 358$ Mpc.  The localization is a probability density map where darker colors indicate higher probability of containing the GW source.  Contours indicate the 50\% and 90\% confidence levels for containing the GW event.}
    \label{fig:S190408an}
\end{figure*}

A candidate GW signal was identified using data from LIGO Livingston Observatory (L1), LIGO Hanford Observatory (L2), and Virgo Observatory (V1) on 2019-04-08 at 18:18:02.288 UTC \citep{gcn24069}.  This event, S190408an, had a $>$99\% probability of being a BBH merger.  The localization covered 387 deg$^2$ at a distance of $1473 \pm 358$ Mpc. S190408an is the first publicly announced astrophysical event in LVC's O3 and the entire localization region resides in the Northern Hemisphere (Figure~\ref{fig:S190408an}), making it a prime target for SAGUARO's first trigger as an important test of the system.

The majority of the localization lies in an area not covered by the CSS survey due to its proximity to the Galactic plane.  As a result, we only triggered three viable fields with template images.  The localization was not immediately observable and the images were taken at a mid-time of $\delta t \approx17.56$~hr after the GW event. As shown in Figure~\ref{fig:S190408an}, these fields covered 8.63 deg$^{2}$ of the 50\% probability region and 13.45 deg$^{2}$ of the 90\% probability region. These observations account for 13.8\% of the 50\% probability, 8.6\% of the 90\% probability, and 7.8\% of the total probability. Poor weather conditions and the low elevation of the fields resulted in a 3$\sigma$ limiting magnitude of 19.8 mag. The details of these observations are summarized in Table~\ref{tab:obs}.

From these 3 fields, 5469 candidates above 5$\sigma$ were detected. No known moving objects or transients were found in the median images after crossmatching these candidates against known moving objects from the MPC and known transients from TNS. Crossmatching to \cite{2014ApJ...788...45T} and \cite{2013ApJS..206....4K} for known active galactic nuclei (AGNs) also found no matches. As generally expected for BBH mergers, no real astrophysical source associated with the GW event was found after candidate vetting.

\subsection{S190425z} \label{sec:S190425z}

\begin{figure*}%
    \centering
    \includegraphics[trim=2.5cm 0cm 3cm 1cm,width=8.5cm]{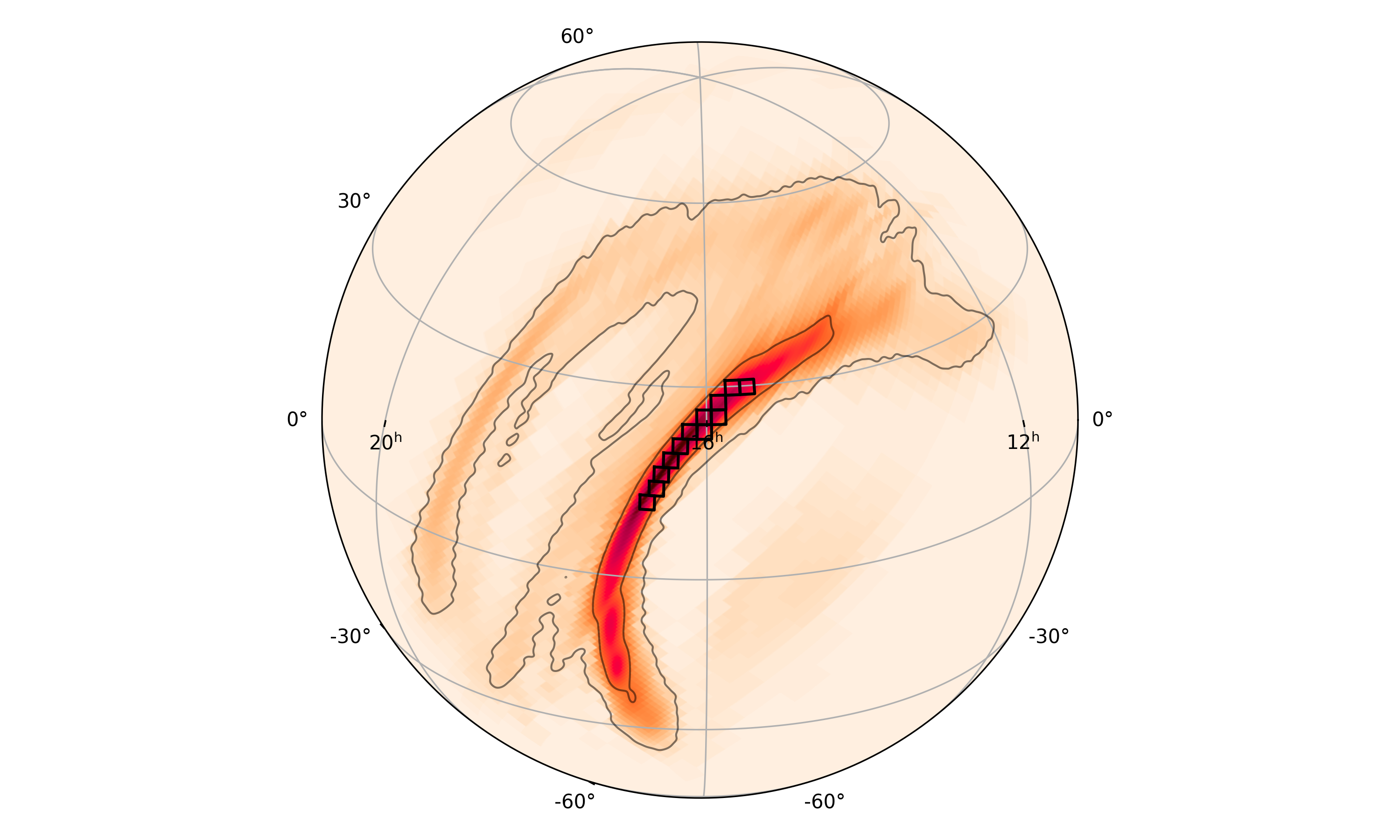} %
    \includegraphics[trim=3cm 0cm 2.5cm 1cm,width=8.5cm]{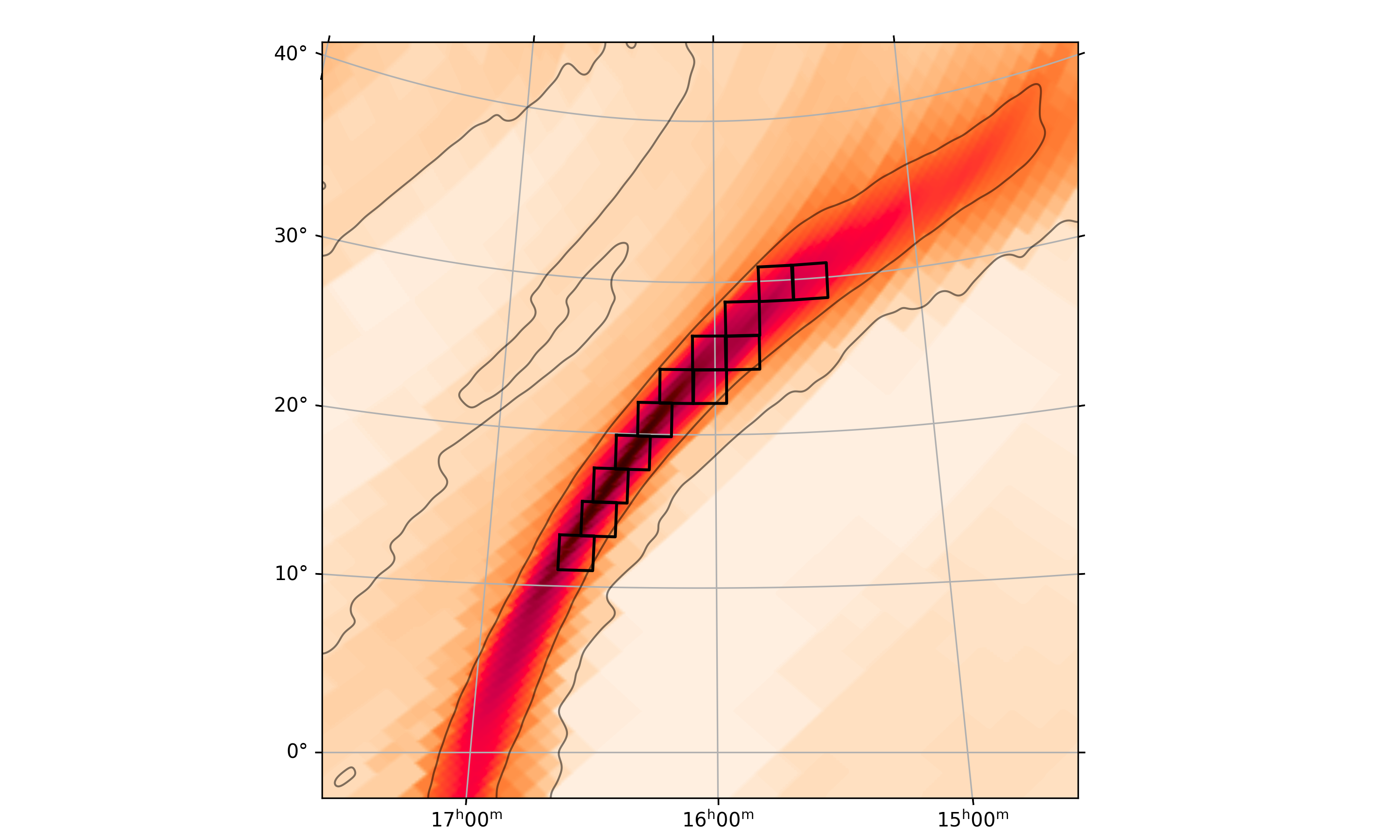} %
    \hfill
    \includegraphics[trim=2.2cm 1cm 1.5cm 0cm,width=12cm]{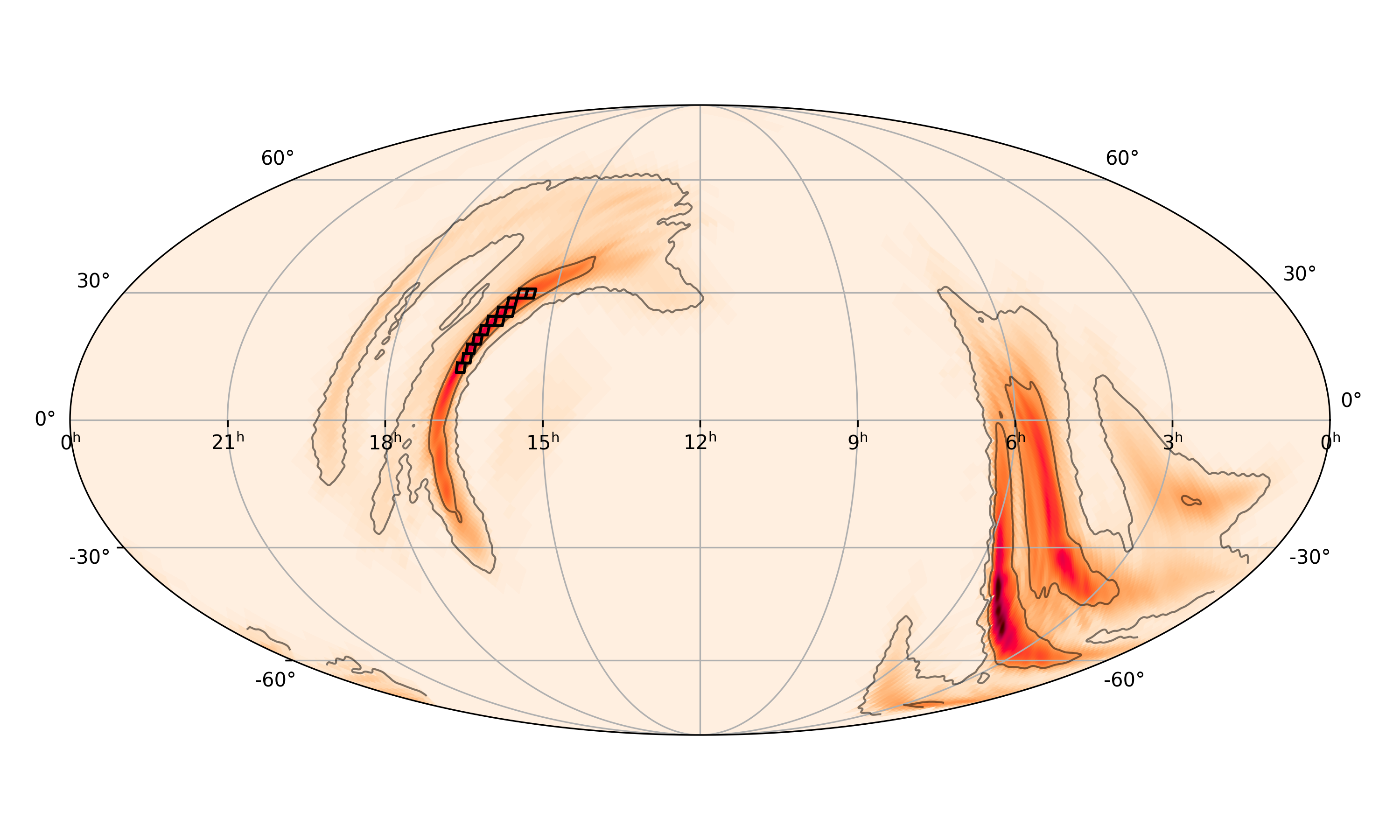} %
    \caption{Same as Figure~\ref{fig:S190408an} for S190425z.  The 90\% localization area covered 7461 deg$^2$ at a distance of $156 \pm 41$ Mpc.  It was concentrated in two regions, one predominately north of the celestial equation near RA$\approx$16h and one mostly south near RA$\approx$5h.  A full Mollweide projection is also shown (bottom) to illustrate the full sky localization.}%
    \label{fig:S190425z}%
\end{figure*}

A candidate GW signal was identified using data from L1 and V1 on 2019-04-25 at 08:18:05.017 UTC \citep{gcn24168}.  This event, S190425z, had a $>$99\% probability of resulting from a BNS merger.  The initial 90\% localization covered 10183 deg$^2$ at a distance of  $155 \pm 45$ Mpc.  The 90\% localization area was later improved to 7461 deg$^2$ at a distance of $156 \pm 41$ Mpc \citep{gcn24228}. It is notable that S190425z was detected as a sub-threshold event in V1, and thus is considered a single detector event, explaining its large localization area with respect to the expected median value for O3.

Once the initial alert was received, the SAGUARO software automatically selected the twelve highest probability fields to observe and placed these into the CSS queue. As Figure~\ref{fig:S190425z} shows, the localization has three distinct regions of high probability.  Two of these regions are mostly visible from the southern hemisphere and one region is mostly visible in the north; the twelve selected fields were naturally in the northern region.  These fields covered 58.5 deg$^{2}$ of the 50\% probability region and 60.0 deg$^{2}$ of the 90\% probability region. These fields account for 6.0\% of the 50\% probability, 3.4\% of the 90\% probability, and 3.0\% of the total probability after the localization was updated (Table~\ref{tab:obs}).

 \begin{figure*}
     \centering
    \includegraphics[width=\textwidth]{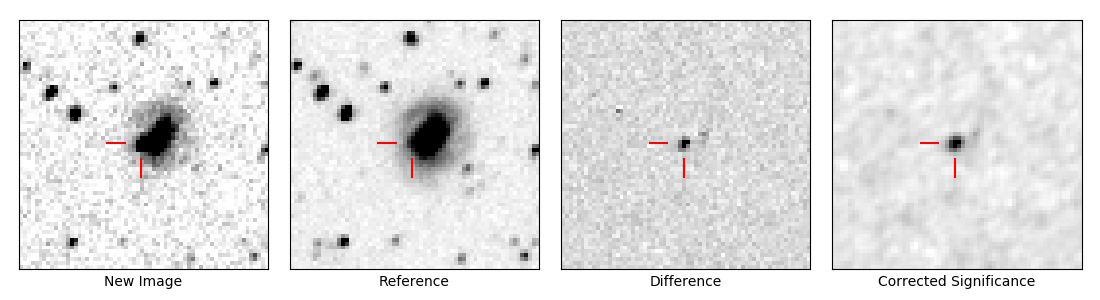}
      \vspace{-0.2in}
    \caption{CSS imaging of AT 2019bln, a known transient found within the data for the S190425z event. Left: The new image created from the median of the 4 CSS images with a mid-time of 2019-04-25 09:35:48.12 UTC and $\delta t \approx1.29$~hr after the GW event. Second: The reference image from the deep templates created with 62 30-sec images, giving a total exposure time of 31 min. Third: The difference image produced by the image subtraction. Right: Corrected significance (Scorr) image showing the significance of the detection in the difference image. AT 2019bln was detected with a SNR of 
    15. 
    }
    \vspace{0.2in}
    \label{fig:sub}
\end{figure*}

The observations of the fields started at $\delta t \approx 1.1$~hr with respect to the GW event, $\sim 0.4$~hr after it was added to the CSS observing queue, with a mid-time of $\delta t \approx 1.37$~hr for all 12 fields.
In favorable observing conditions, we reached a 3$\sigma$ limiting magnitude of 21.3 mag. From the 12 fields observed, 2711 candidates above 5$\sigma$ were detected. After crossmatching these candidates against known moving objects from the MPC, two known moving objects were found to have remained in the median combined images. After crossmatching to the TNS, 6 previously discovered transients were found within the data: AT 2019bln \citep{2019TNSTR.320....1T}, SN 2017frl (\citealt{2017TNSTR.810....1T,2017TNSCR1318....1X}), AT 2019eaj \citep{2019TNSTR.634....1F}, AT 2018cix \citep{2018TNSTR.797....1F}, SN 2019aja \citep{2019TNSTR.175....1S} and SN 2019bzo (\citealt{2019TNSTR.409....1N,2019TNSCR.417....1B}). Figure \ref{fig:sub} shows an example of one such transient found.
Further crossmatching to \cite{2014ApJ...788...45T} and \cite{2013ApJS..206....4K} found 45 known AGNs associated with detections. Four candidates were found within the data: SN 2019eff \citep{2019TNSTR.665....1C}, AT 2019efu \citep{2019TNSTR.665....1C}, AT 2019ech \citep{2019TNSTR.639....1C} and AT 2019fgy \citep{2019TNSTR.777....1C}. Only one candidate, SN 2019eff, was associated with a GLADE galaxy, but at a distance inconsistent with the reported range for the GW event. Spectral classification of SN 2019eff by \citet{2019TNSCR.693....1N} indicated a type IIb supernova.  The remaining candidates were not followed up by any group.

\begin{figure*}
    \centering
   \includegraphics[width=0.49\textwidth,trim={0.0in 0in 0in 0in}]{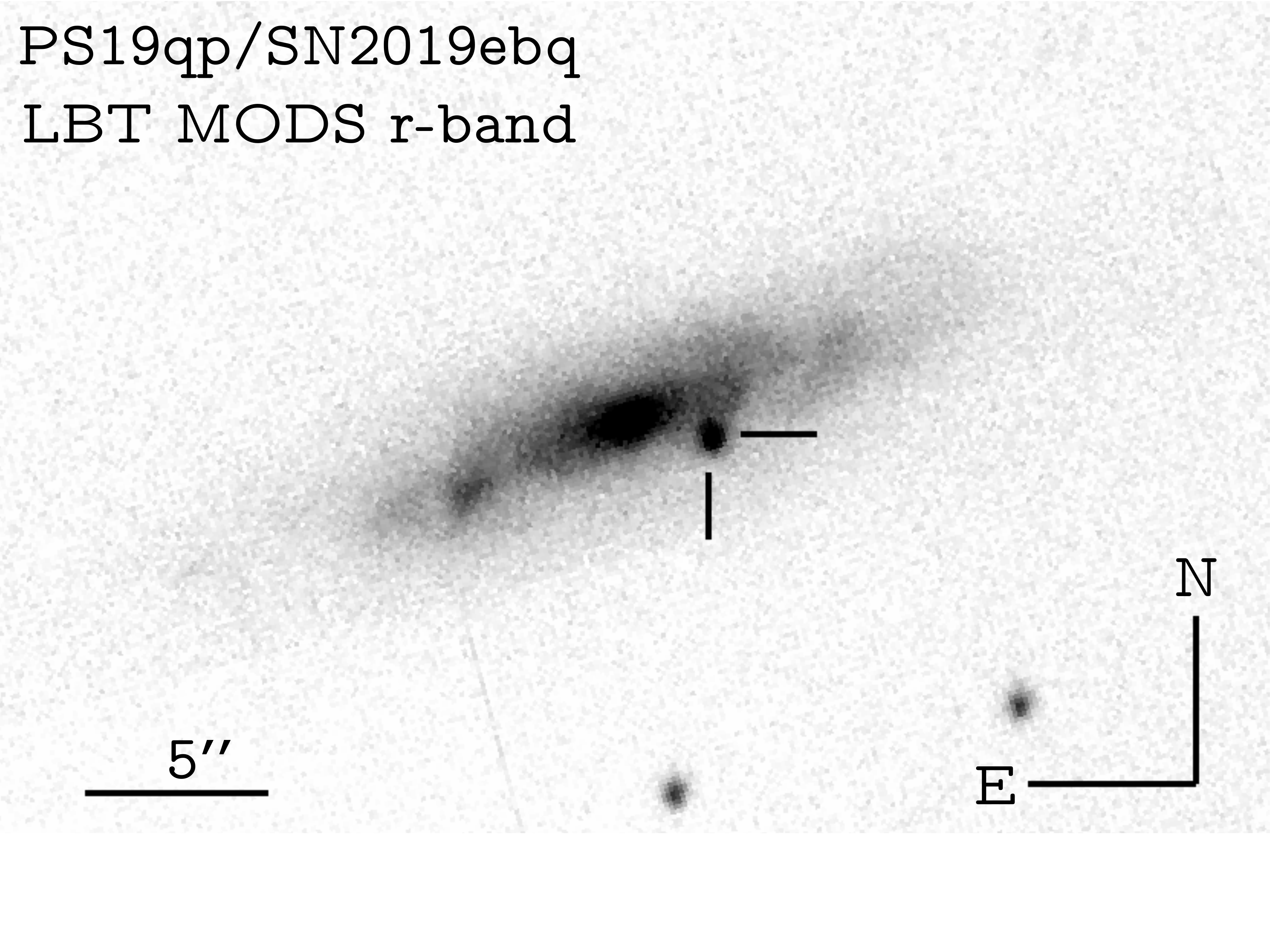}
   \includegraphics[width=0.49\textwidth,clip,trim={0.1in -0.2in 0in 0.5in}]{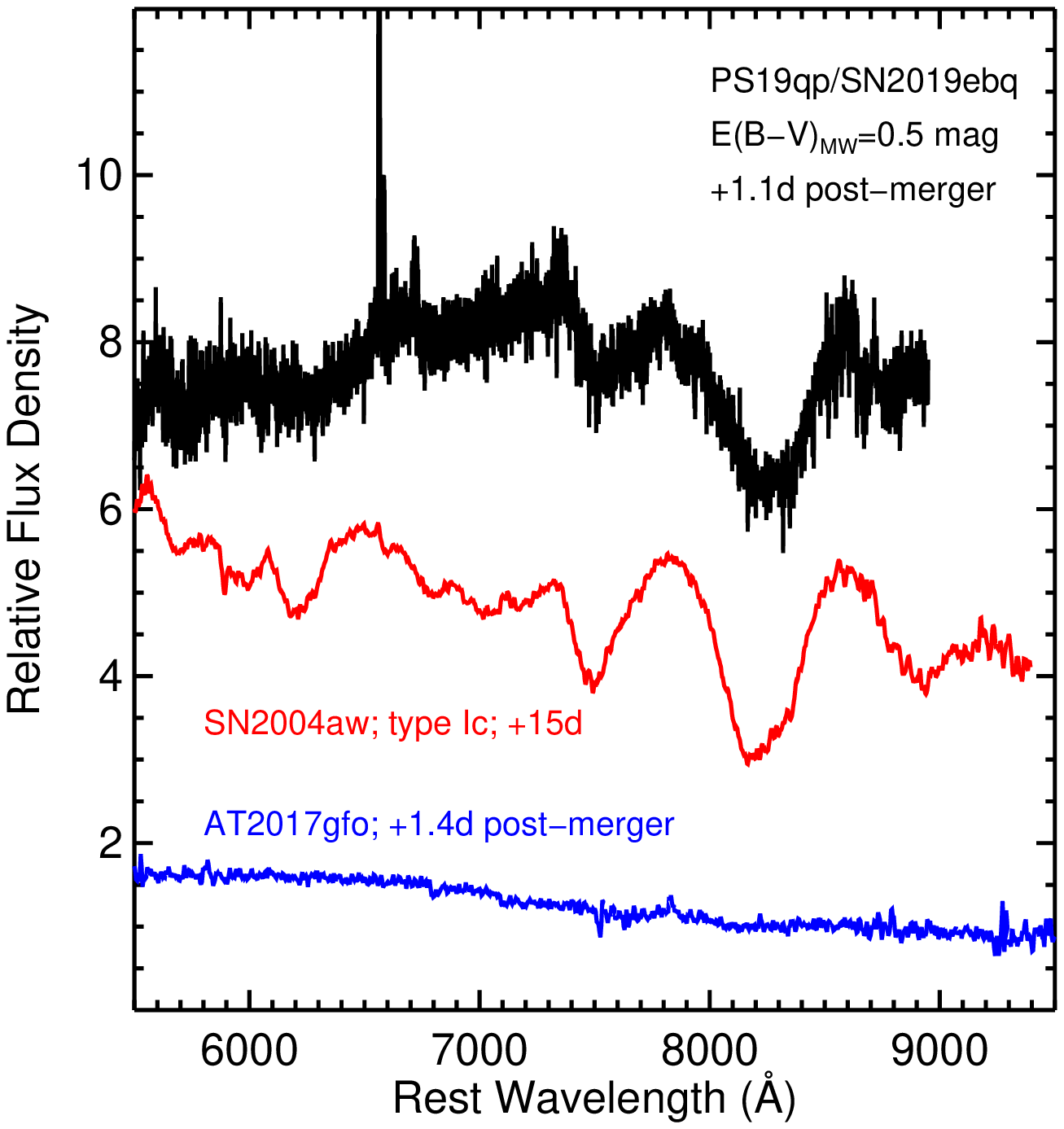}
    \caption{{\it Left:} LBT/MODS $r$-band imaging of PS19qp/SN2019ebq taken  at $\delta t \approx 1.11$~days after the GW trigger of S190425z. The cross-hairs denote the position of the supernova. {\it Right:}  Spectroscopic classification of PS19qp/SN2019ebq using LBT/MODS.  This transient was located in the S190425z localization region, with a host galaxy at distance consistent with the GW trigger.  The comparison spectrum is SN2004aw, a type Ic SN +15 days after maximum light \citep{Taubenberger06}.  For comparison, we show the spectrum of AT2017gfo, the optical counterpart of GW170817, at +1.4d after the GW170817 merger \citep{Valenti2017,Smartt17}.}
    \label{fig:LBTspec}
\end{figure*}

As the follow-up of S190425z developed, 69 GW counterpart candidates were reported in GCNs \citep{Hosseinzadeh19}.  One GW counterpart candidate reported by Pan-STARRS, PS19qp/SN2019ebq \citep[][]{PS19qp_discover}, appeared to be a promising kilonova candidate.  An initial spectrum from the advanced Public ESO Spectroscopic Survey for Transient Objects (PESSTO) displayed a red, featureless continuum and narrow host lines at a redshift $z$=0.037 \citep{PS19qp_pessto}.  This redshift corresponds to a distance consistent with the GW event ($D\approx150$ Mpc), and gave the transient an absolute magnitude of $M_i\approx-16.7$ mag (corrected for Galactic extinction \citealt{Schlafly11}), roughly that of GW170817 at similar epochs. This candidate was not covered by the CSS fields triggered, but we were motivated by the distance, luminosity, and featureless spectrum seen by PESSTO, to trigger ToO imaging and spectroscopy on the LBT starting at $\delta t \approx 1.11$~days after the GW trigger. The transient is well detected in the $60$-sec $r$-band acquisition image (Figure~\ref{fig:LBTspec}). Performing astrometry relative to {\it Gaia} DR2, we measure a location of RA=\ra{17}{01}{18.35}, Dec=$-$\dec{07}{00}{10.5} with a positional uncertainty of $0.060''$ ($1\sigma$). Using standard tasks in IRAF, we measure a brightness for PS19qp/SN2019ebq of $r = 20.44 \pm 0.05$~mag, translating to $M_r\approx-15.59$~mag at $z=0.037$ (or $M_r \approx -16.91$ when accounting for the Galactic extinction in the direction of PS19qp/SN2019ebq of $A_r=1.313$~mag; \citealt{Schlafly11}).

We obtained 2$\times$600s exposures of PS19qp/SN2019ebq at 2019-04-26 11:00  UT using the Multi-Object Double Spectrographs \citep[MODS;][]{MODS} and a 1.2 arcsec slit.  Here we focus on the red-side spectrum, which spanned a usable range of $\sim$5650--9200 \AA, and can be seen in Figure~\ref{fig:LBTspec}; note we have corrected for Milky Way extinction, with a color excess of $E(B-V)$=0.5 mag \citep{Schlafly11}.  Two broad absorption features are apparent, at $\approx$7500\AA~ and 8200\AA, likely due to \ion{O}{1} $\lambda$7774 and the \ion{Ca}{2} IR triplet, respectively -- these spectroscopic signatures are a clear sign that PS19qp/SN2019ebq is a normal supernova. Spectral classification with the Supernova Identification \citep[SNID][]{snid} software package indicates a type Ib/c supernova after maximum light; we plot a +15d spectrum of SN2004aw \citep{Taubenberger06}, a type Ic, in Figure~\ref{fig:LBTspec} to show that this classification is reasonable.  We also show a spectrum of AT2017gfo, the optical counterpart of GW170817, at +1.4d after the GW170817 merger \citep{Valenti2017,Smartt17}, clearly showing that PS19qp/SN2019ebq is an unrelated supernova \citep[see also][]{PS19qp_1,PS19qp_2,PS19qp_3,PS19qp_4,PS19qp_5}.

\subsection{S190426c}

A candidate GW signal was identified using data from L1, L2, and V1 on 2019-04-26 at 15:21:55.337 UTC.  This candidate, S190426c, initially had a 49.3\% probability of resulting from a BNS merger, 23.7\% probability of being a MassGap event, a 12.9\% probability of resulting from a NSBH merger, a 0.0\% probability of being a BBH merger, and a 14.0\% probability of being of terrestrial origin \citep{gcn24237}. The source classification was later significantly revised to a 51.6\% probability of being a NSBH merger, 21.5\% probability of being a MassGap event, and a 12.9\% probability of being a BNS merger with the BBH and terrestrial probabilities unchanged \citep{gcn24411}. The distance of the event was also updated to $377 \pm 100$~Mpc with the 50\% probability region covering 214 deg$^{2}$ and the 90\% probability region covering 1131 deg$^{2}$.

Although a GW170817-like source would not be detectable by our SAGUARO search at this distance, little is known about the expected emission from NSBH mergers. While \cite{2019MNRAS.485.4260S} showed that kilonovae from NSBH mergers should be much dimmer than those from BNS mergers when viewed face-on, the mass ejected is expected to be highly anisotropic, making the viewing angle for these mergers important \citep{2013PhRvD..88d1503K}. Simulations from \cite{2017CQGra..34j4001R} found that NSBH kilonovae reach similar optical peak brightness to BNS kilonovae, while \cite{2014ApJ...780...31T} found that higher ejecta masses could make the kilonovae from NSBH events more luminous than with BNS mergers. With such a wide range of predictions for the emission from NSBH mergers, follow-up of these sources is crucial to place the first observational constraints on models.

\begin{figure*}
    \centering
   \includegraphics[trim=2.5cm 0cm 3cm 0cm,width=8.5cm]{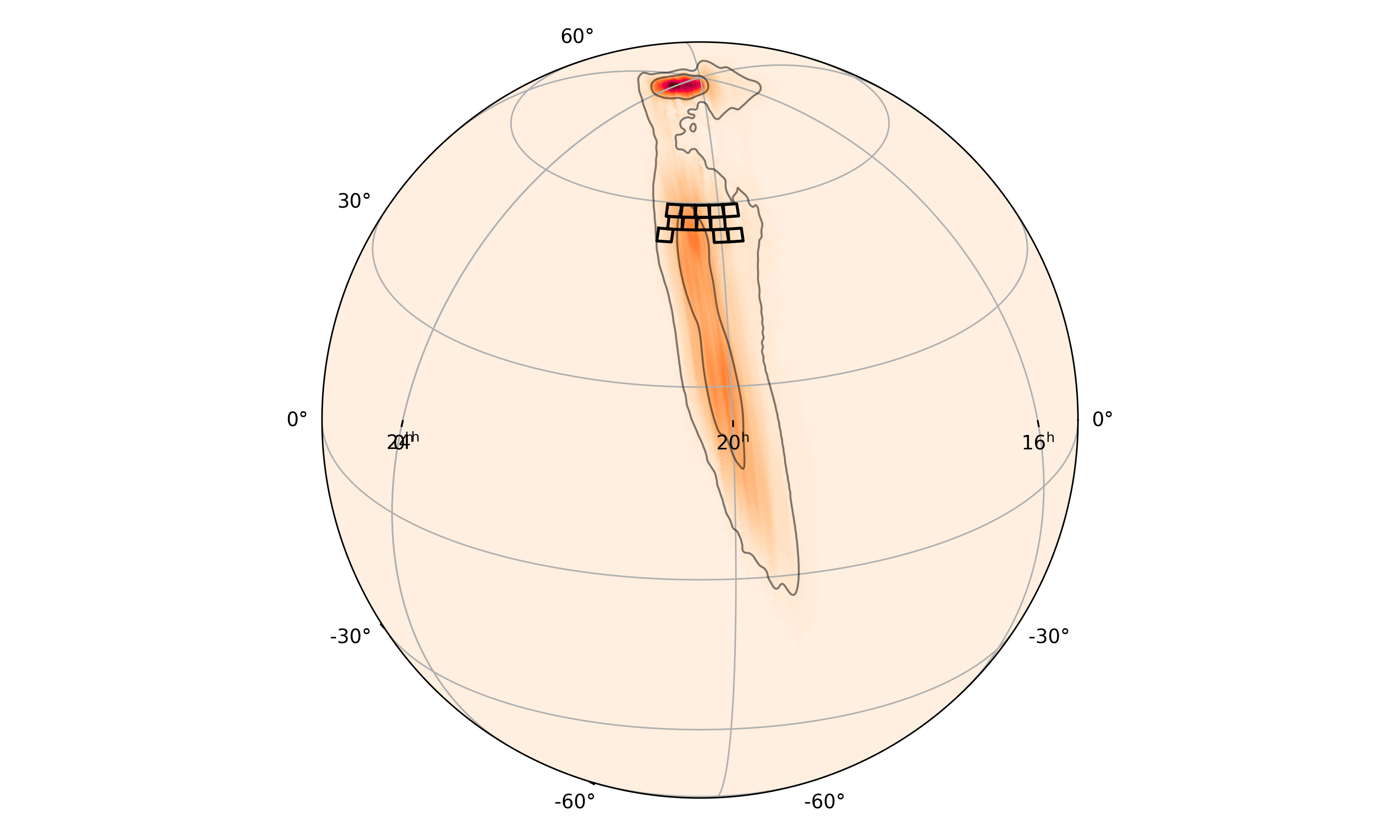} %
    
    \includegraphics[trim=3cm 0cm 2.5cm 0cm,width=8.5cm]{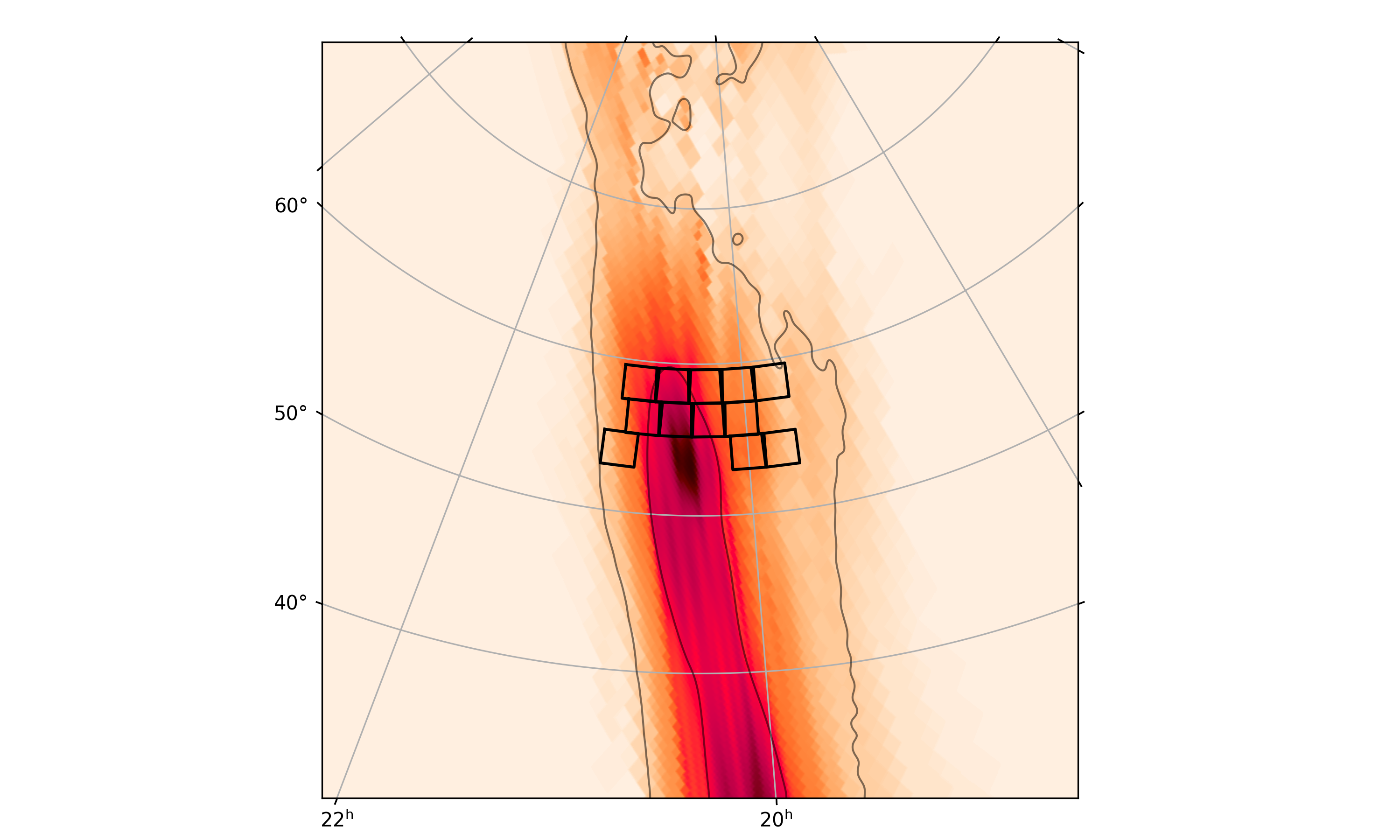} 
    \caption{Same as Figure~\ref{fig:S190408an} for S190426c, showing the updated localization.  This event occured at a distance of $377 \pm 100$~Mpc.  The 50\% probability region covered 214 deg$^{2}$ and the 90\% probability region covered 1131 deg$^{2}$.  The twelve fields selected to be observed were limited by our +60 deg declination limit and lack of coverage near the Galactic plane.}
    \label{fig:S190426c}
\end{figure*}

Once the GW alert was received, the twelve fields shown in Figure~\ref{fig:S190426c} were selected to be observed.  The highest probability regions of the localization were either north of our +60 deg declination limit or fell in an area near the Galactic plane that is not covered by the CSS survey due to crowded fields.  As a result, we canceled the initial trigger and triggered the telescope (selecting the same fields from the initial localization) after the localization map was updated \citep{gcn24277}. These fields were observed at a mid-time of $\delta t \approx 1.74$~days, covering 13.4 deg$^{2}$ of the 50\% probability region and 58.9 deg$^{2}$ of the 90\% probability region. These fields account for 4.3\% of the 50\% probability, 5.6\% of the 90\% probability, and  5.1\% of the total probability after the localization was updated (e.g., \citealt{gcn24277}). A slightly lower elevation and brighter sky resulted in a 3$\sigma$ limiting magnitude of 20.8 mag.

From the triggered fields, 11307 candidates above 5$\sigma$ were detected. No known moving objects, transients or AGNs were found in the median images after crossmatching these candidates against known moving objects from the MPC, known transients from TNS, and known AGNs \citep{2014ApJ...788...45T,2013ApJS..206....4K}. One candidate was found within the data (AT 2019eij, \citealt{2019TNSTR.679....1D}) and was not associated with a GLADE galaxy.

\section{Summary and Future Prospects} \label{sec:summary}
We introduce the SAGUARO GW optical counterpart search strategy, where we tile the highest probability regions of the localization with the Steward Observatory Mt. Lemmon 1.5m telescope and its 5 deg$^2$ imager.  With this facility we obtain observations covering 60 or 120 deg$^2$ down to $\sim21.3$ mag within $\delta t \lesssim 30$~min of receiving the GW alert if it is observable.  We also present the SAGUARO software suite that we utilize to trigger the observations, process the data, and display the candidates for vetting.  The robustness of this system allows us to start tiling the localization within minutes of receiving the GW alert and produce lists of candidates within hours.

We discuss SAGUARO's first results from the optical counterpart searches following three GW events.  For S190408an we tiled 15 deg$^2$ down to a limit of 19.8 mag.  For S190425z we tiled 60 deg$^2$ down to a limit of 21.3 mag.  For S190426c we tiled 60 deg$^2$ down to a limit of 20.8 mag. Across all 3 events we found a total of 19487 candidates. Crossmatching these candidates to catalogs of moving objects, previously reported transients, and AGNs, we were able to detect 2 moving objects, 6 previously found transient and 45 AGN. After visual inspection, we found 5 real candidates for the three events. One candidate had a match to a GLADE galaxy, but at the wrong distance and was subsequently classified as a supernova.  The remaining candidates were not associated with GLADE galaxies and were not followed up.  We also spectroscopically classified a candidate reported by the community within $\approx 1.1$~days of the GW event.

Given that the median expected localizations for BNS mergers prior to the LVC's O4 is $\approx 120-180$~deg$^2$ \citep{lvc_loc}, it is useful to increase the FOV of our searches in the near future. To this end, we anticipate bringing online additional resources that will complement the existing optical counterpart search of SAGUARO, including the CSS-run Schmidt 0.7m telescope on Mt. Bigelow in Arizona which uses a 10.5K x 10.5K CCD with a 20 deg$^{2}$ FOV. With this FOV, these BNS merger localizations could be covered in $<10$~pointings. Reaching typical depths of V$\sim$19.5 in 30s exposures, this would allow searches for BNS out to $\sim$ 100 Mpc.

Beyond O3, the Kamioka Gravitational Wave Detector (KAGRA; \citealt{KAGRA}) is expected to join the LVC. LIGO is also expected to undergo a significant upgrade to ``A+''\footnote{https://dcc.ligo.org/LIGO-T1800133/public}. Coupled together, the event rates as well as the fraction of well-localized BNS mergers will increase, with predicted median localizations for BNS mergers of only $\approx 9-12$~deg$^{2}$ \citep{lvc_loc}. In order to match the capabilities of future GW detectors and improvements, we will continue to explore opportunities on larger aperture optical and NIR facilities which will serve as valuable resources for distant, well-localized events.

SAGUARO is able to discover and characterize kilonovae of comparable luminosity to GW170817 out to 200~Mpc on timescales of $\delta t \lesssim 1$~day, with monitoring capabilities on the timescale of weeks.  The ability of software systems to automatically trigger wide-field telescopes, combined with the real-time processing of difference images, is crucial for the rapid detection of kilonova candidates.  In this role, SAGUARO represents a significant addition to the search for optical counterparts to GW events.

\acknowledgments

SAGUARO is supported by the National Science Foundation under Award Nos.\ AST-1909358 and AST-1908972. Research by DJS is also supported by NSF grants AST-1821987, AST-1821967, AST-1813708, and AST-1813466. Research by K.P. and W.F. is also supported by NSF Award No.\ AST-1814782. Research by JCW is supported by NSF AST-1813825.
JS acknowledges support from the Packard Foundation. EP and AR acknowledge funding from the GRAvitational Wave Inaf TeAm (GRAWITA).  The UCSC team is supported in part by NASA grant NNG17PX03C, NSF grant AST-1518052, the Gordon \& Betty Moore Foundation, the Heising-Simons Foundation, and by a fellowship from the David and Lucile Packard Foundation to R.J.F. A.R. acknowledges support from Premiale LBT 2013. This work was partially performed at the Aspen Center for Physics, which is supported by National Science Foundation grant PHY-1607611. This research was supported in part by the National Science Foundation under Grant No. NSF PHY-1748958.  A.C. acknowledges support from the National Science Foundation under CAREER award \#1455090. A.I.Z. acknowledges support from Data7: UA's Data Science Institute. MRD acknowledges support from the Dunlap Institute at the University of Toronto and the Canadian Institute for Advanced Research (CIFAR).
The operation of the facilities of Steward Observatory are supported in part by the state of Arizona.

The LBT is an international collaboration among institutions in the United States, Italy and Germany. LBT Corporation partners are: The University of Arizona on behalf of the Arizona Board of Regents; Istituto Nazionale di Astrofisica, Italy; LBT Beteiligungsgesellschaft, Germany, representing the Max-Planck Society, The Leibniz Institute for Astrophysics Potsdam, and Heidelberg University; The Ohio State University, and The Research Corporation, on behalf of The University of Notre Dame, University of Minnesota and University of Virginia.

This research has made use of data and/or services provided by the International Astronomical Union's Minor Planet Center.

\vspace{5mm}
\facilities{LBT Consortium Large Binocular Telescope (LBT) at Mount Graham International Observatory, Steward Observatory 1.5m (60inch) Telescope (part of the Catalina Sky Survey; CSS) at Mount Lemmon Observing Facility (MLOF), University of Arizona/Smithsonian Institution 6.5m MMT Telescope at Fred Lawrence Whipple Observatory (FLWO), Steward Observatory 2.3m (90inch) Bart Bok Telescope at Kitt Peak National Observatory (KPNO), California Association for Research in Astronomy 10m W.M. Keck I Telescope at Mauna Kea Observatory, California Association for Research in Astronomy 10m W.M. Keck II Telescope at Mauna Kea Observatory, Steward Observatory 1.54m (61inch) Kuiper Telescope (formerly NASA telescope) at Catalina Station, Vatican Observatory Research Group 1.8m Alice P. Lennon Vatican Advanced Technology Telescope (VATT) at Mount Graham International Observatory, Carnegie Institution for Science (CIS) 6.5m Landon Clay Telescope at Las Campanas Observatory (LCO), Carnegie Institution for Science (CIS) 6.5m Walter Baade Telescope at Las Campanas Observatory (LCO), Steward Observatory 0.7m (28inch) Schmidt Telescope (part of the Catalina Sky Survey; CSS) at Mount Lemmon Observing Facility (MLOF)}

\software{
astropy \citep{2013A&A...558A..33A,astropy},   
The IDL Astronomy User's Library \citep{IDLforever}, SCAMP \citep{scamp,scamp2}, SWarp \citep{swarp}, IRAF \citep{iraf1,iraf2},
          SExtractor \citep{Bertin1996}, ZOGY (\url{https://github.com/pmvreeswijk/ZOGY})
          }

\bibliographystyle{aasjournal}
\bibliography{biblio}

\end{document}